\documentclass[aps,pra,showpacs,twocolumn]{revtex4}
\usepackage{amsmath}
\usepackage{graphicx}
\usepackage{float}
\usepackage[ansinew]{inputenc}
\usepackage{array}
\usepackage{color}
\usepackage{amsxtra}
\usepackage{amstext}
\usepackage{amssymb}
\usepackage{latexsym}
\usepackage{gensymb}
\usepackage{dsfont}
\usepackage{braket}
\begin{document}

\title{Analytical solvability of the two-axis countertwisting spin squeezing Hamiltonian}

\author{M. Bhattacharya}
\affiliation{School of Physics and Astronomy, Rochester Institute of Technology, 84 Lomb Memorial Drive,
Rochester, NY 14623, USA}

\date{\today}
\begin{abstract}
There is currently much interest in the two-axis countertwisting spin squeezing Hamiltonian suggested
originally by Kitagawa and Ueda, since it is useful for interferometry and metrology. No analytical
solution valid for arbitrary spin values seems to be available. In this article we systematically
consider the issue of the analytical solvability of this Hamiltonian for various specific spin values.
We show that the spin squeezing dynamics can be considered to be analytically solved for angular momentum 
values upto $21/2$, i.e. for $21$ spin half particles. We also identify the properties of the system 
responsible for yielding analytic solutions for much higher spin values than based on naive expectations. 
Our work is relevant for analytic characterization of squeezing experiments with low spin values, and
semi-analytic modeling of higher values of spins.
\end{abstract}

\maketitle
\section{Introduction}
The two-axis countertwisting spin-squeezing Hamiltonian $(\hbar=1)$
\begin{equation}
\label{eq:HTA}
H_{TA}=\frac{\chi}{2i}\left(J_{+}^{2}-J_{-}^{2}\right),
\end{equation}
was originally proposed by Kitagawa and Ueda \cite{Kitagawa1991,Kitagawa1993}. In Eq.~(\ref{eq:HTA}) $J_{\pm}$
are the angular momentum raising and lowering operators \cite{SakuraiBook}, $i=\sqrt{-1}$ and $\chi$ is a
constant. The quantity $J$ may refer to a real angular momentum or a pseudospin $J=N/2$ representing the
collective squeezing of $N$ spin half systems \cite{Ma2011}. The Hamiltonian $H_{TA}$ yields maximal
squeezing, with a squeezing angle independent of system size or evolution time. Experimental implementation
has not yet been achieved, while a number of theoretical proposals have been put forward
\cite{Berry2002,Liu2011,Nemoto2014,Opatrny2014,Huang2015,Das2015,Li2015}.

A general analytic solution to the Hamiltonian for arbitrary angular momentum does not seem to be available
\cite{Jafarpour2008,Pathak2008}. Solutions to the dynamics for low values of spin are of interest to
experiments with trapped ions \cite{Meyer2001} and quantum magnets \cite{Toth2009}, for example. In 
Ref.~\cite{Ma2011} a bound of spin $3/2$ was stated as the maximum angular momentum value for which 
analytic solutions can be found. We presume this was based on the fact that the Hamiltonian matrix in that 
case is of dimension $2\times 3/2+1=4,$ leading to a characteristic polynomial of quartic order, which is 
the highest degree for which algebraic solutions can generally be found \cite{StewartBook}. In 
Ref.~\cite{Jafarpour2008} the existence of analytical solutions (with the additional presence of an external 
field) up to spin $10$ was reported and explicit expressions were provided for spin $1$. No explanation was 
given of the unexpected fact that solutions for spins much larger than $3/2$ were found, nor was the maximum 
value of spin for which analytical solutions could be found supplied.

In the present article we systematically analyze the question of analytical solvability of the two-axis
countertwisting spin squeezing Hamiltonian of Eq.~(\ref{eq:HTA}). We extend the bound for solvability to spin 
$21/2$, i.e. $21$ spin half particles (we note that in Ref.~\cite{Jafarpour2008} spin $10$ corresponds to $10$
spin half particles). We point out that critical roles in the solvability of the model for large spin values are 
played by the chiral symmetry and sparsity of the Hamiltonian matrix. Our approach requires only the use of 
matrix representations of the angular momentum operators and the evaluation the time evolution operator 
\cite{Zela2014}. Our results may be useful for experiments with small number of spins.

\section{Some properties of $H_{TA}$}
In this section we go over some properties of the Hamiltonian $H_{TA}$ of Eq.~(\ref{eq:HTA}) which are relevant
to the discussion of analytical solvability. First, it can be seen readily by using
\begin{equation}
J_{+}=J_{-}^{\dagger},
\end{equation}
that $H_{TA}$ is Hermitian, implying that its eigenvalues are real.

Further, by using
\begin{equation}
J_{\pm}=J_{x}\pm i J_{y},
\end{equation}
we can rewrite
\begin{equation}
H_{TA}=\chi\left(J_{x}J_{y}+J_{y}J_{x}\right).
\end{equation}
Now let us consider a rotation by the angle $\pi$ about the $J_{y}$ axis. This rotation leaves $J_{y}$
unaffected, but reverses the sign of $J_{x}$, i.e.
\begin{equation}
e^{-i\pi J_{y}}H_{TA}e^{i\pi J_{y}}=-H_{TA},
\end{equation}
which can be written as the anticommutation relation
\begin{equation}
\label{eq:AntiC}
\{H_{TA},e^{i\pi J_{y}}\}=0.
\end{equation}
This relation implies that $e^{i\pi J_{y}}$ is a chiral symmetry of $H_{TA}$ \cite{Mishkat2014}. In
practical terms, the implication of the anticommutation is that eigenvalues of $H_{TA}$ occur as signed pairs
$\pm \lambda_{1},\pm \lambda_{2},\ldots$ (A brief proof is provided in the Appendix for the reader's convenience).
Therefore if $H_{TA}$ can be represented by an even dimensional matrix, then the characteristic polynomial of
$H_{TA}$
\begin{equation}
P(\lambda)=\left|H_{TA}-\lambda I\right|=0,
\end{equation}
is even in $\lambda$, where $I$ is the unit matrix of the same dimension as $H_{TA}$. If instead $H_{TA}$
is of odd dimension, then its characteristic polynomial is $\lambda$ times a polynomial even in $\lambda$.
In this case there is necessarily a zero eigenvalue, and all other eigenvalues are signed pairs. Specific
examples will be given below. As will be verified with these examples, the property of chiral symmetry
contributes to giving a simple form to the characteristic polynomials for even large spins, and making them
analytically solvable. For completeness we note that the Hamiltonian considered by the authors of
Ref.~\cite{Jafarpour2008}
\begin{equation}
H_{f}=H_{TA}+\Omega J_{z},
\end{equation}
where $\Omega$ represents an external field along the $z$ axis, also possesses the chiral symmetry indicated
above. This partly explains the solvability of the squeezing model $H_{f}$ analytically for up to spin $10$.

Finally, by using the matrix elements in the basis $|j,m\rangle$ where $J_{z}$ is diagonal
\begin{equation}
\langle j,m'|J_{\pm}|j,m\rangle = \sqrt{j(j+1)-m(m'\pm 1)},
\end{equation}
it can be readily verified that $H_{TA}$ is a rather sparse matrix, i.e. most of its elements are zero. This
feature also contributes to simplifying the form of the characteristic polynomial.

\section{The time evolution operator}
We now consider the time evolution operator
\begin{equation}
\label{eq:UTA}
U_{TA}=e^{-iH_{TA}t}.
\end{equation}
Since $U_{TA}$ determines the spin dynamics completely, the model is analytically solvable if a matrix
representation for $U_{TA}$ can be found, with all entries determined analytically. A straightforward way
to implement this is to diagonalize $H_{TA}$. If the eigenvalues of $H_{TA}$ can be found analytically, the
diagonal and analytic form of $U_{TA}$ follows. However, the calculation of expectation values then requires
the relevant initial state to be rotated by the unitary matrix that diagonalizes $H_{TA}.$ As these matrices
can be quite cumbersome and require the determination and careful handling of the eigenvectors of $H_{TA},$
we follow instead an equivalent procedure that only deals with the eigenvalues of $H_{TA},$ but avoids
referring to its eigenvectors.

Our approach is to expand the right hand side of Eq.~(\ref{eq:UTA}) in a Taylor series. The termination of that
series is actually guaranteed since $H_{TA}$ is a finite dimensional matrix. This guarantee comes from the
Cayley-Hamilton theorem, which states that every square matrix satisfies its own characteristic equation, which
in turn implies that any power of the $H_{TA}$ can be expressed in terms of the matrix powers
$(H_{TA})^{0},(H_{TA})^{1},\ldots, (H_{TA})^{2J},$ where $J$ is the associated spin. The entries of the
terminating matrix representation of $U_{TA}$, it turns out, can be written as functions of the eigenvalues of
$H_{TA}$ \cite{Zela2014}, see below. Therefore, if the eigenvalues can be calculated analytically, then the spin
dynamics can be found exactly.

For reference, we quote the explicit expression for $U_{TA}$ in the case where $H_{TA}$ possesses nondegenerate 
eigenvalues $\lambda_{k}, k=1,2,\ldots$ (the degenerate case can be handled via some straightforward
modifications) \cite{Zela2014})
\begin{equation}
\label{eq:UTAZela}
U_{TA}=\displaystyle \sum_{k = 1}^{2j+1}e^{-i\lambda_{k}t}
\displaystyle\prod_{{\substack{n=1 \\ n\neq k}}}^{2j+1} \left(\frac{H_{TA}-\lambda_{n}}{\lambda_{k}-\lambda_{n}}\right).
\end{equation}

\section{Results}
\subsection{Solvability}
In Table 1. we show for various spins the characteristic polynomial $P(\lambda)$ of $H_{AT}$. We make some
comments on the entries in this table. For $J=1/2$, the eigenvalues are both zero, and it can be verified
that
\begin{equation}
H_{TA}(J=1/2)=
\left(
  \begin{array}{cc}
    0 & 0 \\
    0 & 0 \\
  \end{array}
\right),
\end{equation}
consistent with the observation, first made by Kitagawa and Ueda, that a single spin half particle cannot be
squeezed \cite{Kitagawa1993}.

Generally for half-integer spin, $P(\lambda)$ is even in $\lambda$ and some of the eigenvalues are degenerate.
In Table 1, we have indicated which spin values lead to degeneracy in the eigenvalues of $H_{TA}$, so that the
appropriate procedure may be used for obtaining $U_{TA}$. Generally, if any factor in the characteristic
polynomial repeats, then there is degeneracy in the energy spectrum. To make this identification rigorous, we
have calculated the discriminant of $P(\lambda)$, which returns a zero value if degeneracy is present. The
combination of chiral symmetry and matrix sparseness leads to rather compact expressions for the spin
half-integer $P(\lambda)$'s. Indeed it is remarkable that $P(\lambda)$ for $J=21/2$ fits on a single line.
For integer spin, $P(\lambda)$ is a polynomial even in $\lambda$, times a single factor of $\lambda$. There
seems to be no degeneracy in general and the polynomials are less compact in form than in the half-integer
spin case.

Upto $J=9/2$ it is evident that the roots of the polynomials can be found analytically, since the factors
are of degree $4$ or less, and solvability by radicals is guaranteed by the Abel-Ruffini theorem
\cite{StewartBook}. While spins $5$ to $13/2$ have factors sextic in $\lambda$, since these polynomials
are even in $\lambda$, they can be thought of as being cubics in $\lambda^{2},$ which can be solved
analytically. The same reasoning applies to the octic factors in the polynomials for spins $7$ to $17/2.$
Finally, spin $9$ to $21/2$ contain factors of degree $10$, which can be considered as being polynomials
of degree $5$ in $\lambda^{2}.$ While the roots of such factors cannot be found algebraically, they can
be stated in terms of hypergeometric functions \cite{King1991}. However, for $J=11$, there is a factor of
degree $12$ in $\lambda$ (i.e. of degree $6$ in $\lambda^{2}$). While the roots of polynomials of degree $6$
and higher can be found in terms of modular functions (for example), they involve transcendental functions,
and we will consider them not to be of closed form and therefore analytically unsolvable \cite{BoydBook}. 
We note that for $J>21/2$ the polynomial roots can be found numerically and inserted in Eq.~(\ref{eq:UTAZela}), 
thus yielding a semianalytic solution for any spin value.

\subsection{Spin squeezing}
\begin{figure} 
\includegraphics[width=0.4\textwidth]{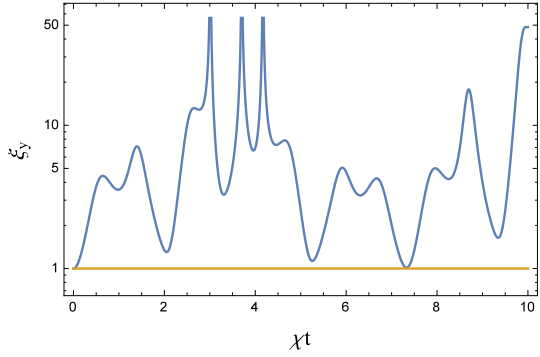}
\caption{(Color online) The squeezing parameter $\xi_{y}$ [Eq.~(\ref{eq:WineSqueeze1})] as a function of the dimensionless
time $\chi t$ for $J=2$.}
\label{fig:P1}
\end{figure}

\begin{figure}
\includegraphics[width=0.4\textwidth]{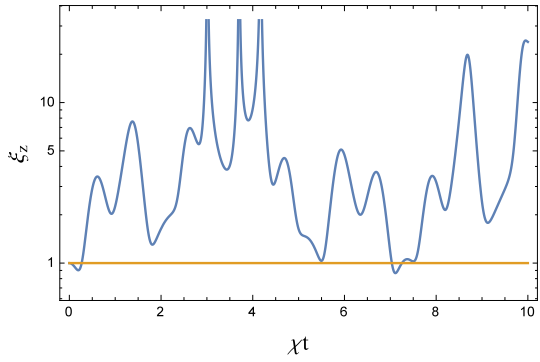}
\caption{(Color online) The squeezing parameter $\xi_{z}$ [Eq.~(\ref{eq:WineSqueeze2})]  as a function of the dimensionless
time $\chi t$ for $J=2$.}
\label{fig:P2}
\end{figure}

To compactly illustrate our results, we present the details for $J=2$ ($4$ spin half particles), a case for which
there seem to be no explicit results in the literature. In this instance, the matrix representations are given by
\begin{equation}
J_{+}=
\left(
  \begin{array}{ccccc}
    0 & 2 & 0 & 0 & 0 \\
    0 & 0 & \sqrt{6} & 0 & 0 \\
    0 & 0 & 0 & \sqrt{6} & 0 \\
    0 & 0 & 0 & 0 & 2 \\
    0 & 0 & 0 & 0 & 0 \\
  \end{array}
\right)
\end{equation}
$J_{-}=J_{+}^{\dagger}$, and
\begin{equation}
H_{TA}=i\chi
\left(
  \begin{array}{ccccc}
    0 & 0 & -\sqrt{6} & 0 & 0 \\
    0 & 0 & 0 & -3 & 0 \\
  \sqrt{6} & 0 & 0 & 0 & -\sqrt{6} \\
    0 & 3 & 0 & 0 & 0 \\
    0 & 0 & \sqrt{6} & 0 & 0 \\
  \end{array}
\right),
\end{equation}
The eigenvalues of $H_{TA}$ can be read off from Table 1. as $0,\pm 3,\pm 2\sqrt{3}$.
The eigenvectors are $(-1, 0, i\sqrt{2}, 0,1)/2, (-1, 0, -i\sqrt{2}, 0, 1)/2, (0, i, 0, 1, 0)/\sqrt{2}$,
$(0, -i, 0, 1, 0)/\sqrt{2}$, and $(1, 0, 0, 0, 1)/\sqrt{2}$ in no specific order. The time evolution 
operator is
\begin{widetext}
\begin{equation}
U_{TA}=
\left(
  \begin{array}{ccccc}
    \cos^{2}(\sqrt{3}\chi t)& 0 & -\sin(2\sqrt{3}\chi t)/2 & 0 & \sin^{2}(\sqrt{3}\chi t)  \\
    0 & \cos(3\chi t) & 0 & -\sin(3\chi t) & 0 \\
  \sin(2\sqrt{3}\chi t)/2 & 0 & \cos(2\sqrt{3}\chi t) & 0 & -\sin(2\sqrt{3}\chi t)/2 \\
    0 & \sin(3\chi t) & 0 & \cos(3\chi t) & 0 \\
    \sin^{2}(\sqrt{3}\chi t)  & 0 & \sin(2\sqrt{3}\chi t)/2 & 0 & \cos^{2}(\sqrt{3}\chi t) \\
  \end{array}
\right).
\end{equation}
\end{widetext}
The time evolution of any operator $\mathcal{O}$ is given by $\mathcal{O}(t) = U_{TA}^{-1}\mathcal{O}U_{TA}.$
Using this relation we can find the time-evolved quantities $J_{y}(t), J_{y}^{2}(t),$ etc., and variances
such as
\begin{equation}
\left(\Delta J_{y,}(t)\right)^{2}=\langle J_{y,}^{2}(t)\rangle-\langle J_{y}(t)\rangle^{2},
\end{equation}
etc. Starting from the initial state
\begin{equation}
\ket{i}=e^{i\pi J_{y}/2}\ket{j=2,m=2}=\frac{1}{2}\ket{\frac{1}{2},1,\sqrt{\frac{3}{2}},1,\frac{1}{2}},
\end{equation}
i.e. the stretched state along $z$ rotated by $90^\circ$ about the $y$ axis \cite{Kitagawa1993}, we find the
squeezing parameters following Wineland et al. \cite{Wineland1992,Wineland1994}
\begin{eqnarray}
\label{eq:WineSqueeze1}
\xi_{y}&=&\sqrt{4}\frac{\langle \left(\Delta J_{y}(t)\right)\rangle}{\left|\langle J_{x}(t)\rangle \right|},\nonumber\\
&=&\frac{\sqrt{2}(17-6\cos6\chi t-6\cos2\sqrt{3}\chi t+3\cos4\sqrt{3}\chi t)^{1/2}}
{\left|\cos3\chi t(1+3\cos2\sqrt{3}\chi t)+\sqrt{3}\sin3\chi t\sin2\sqrt{3}\chi t\right|},\nonumber\\
\end{eqnarray}
and
\begin{widetext}
\begin{table}[hb]
\renewcommand*{\arraystretch}{1.3}
\caption{Characteristic Polynomials of $H_{TA}/\chi$} 
\centering
\begin{tabular}{c c c}
\hline\hline
\noalign{\smallskip}
$J$ & $P(\lambda)$ & $\mathrm{Degenerate}$ \\ [0.5ex]
\hline\hline
\noalign{\smallskip}
$1/2$ & $\lambda^{2}$ & Yes \\
\hline
$1$ & $\lambda (1-\lambda^{2})$ & No \\
\hline
$3/2$ & $(\lambda^{2}-3)^{2}$ & Yes \\
\hline
$2$& $-\lambda (\lambda^{2}-3)(\lambda^{2}-12)$ &  No \\
\hline
$5/2$& $\lambda^{2}(\lambda^{2}-28)^{2}$ & Yes \\
\hline
$3$& $-\lambda (\lambda^2-60) (\lambda^2-6 \lambda-15) (\lambda^2+6 \lambda-15)$ & No\\
\hline
$7/2$& $(\lambda^{4} - 126 \lambda^{2}+945)^{2}$ & Yes \\
\hline
$4$ & $-\lambda (\lambda^{2}-28)(\lambda^{2}-208)$ &  \\
    & $\times (\lambda^{2}+10\lambda-63)(\lambda^{2}-10\lambda-63)$ & No \\
\hline
$9/2$ & $\lambda^{2}(\lambda^{4}-396\lambda^{2}+19008)^{2}$ & Yes  \\
\hline
$5$ & $-\lambda (\lambda^{2}-108)(\lambda^{2}-528)$ &  \\
    & $\times (\lambda^{6}-651\lambda^{4}+65619\lambda^{2}-455625)$ & No \\
\hline
$11/2$& $(\lambda^{6} - 1001 \lambda^{4}+172315\lambda^{2}-2338875)^{2}$ & Yes \\
\hline
$6$ & $-\lambda (\lambda^{2}-336)(\lambda^{4}-1176\lambda^{2}+55440)$ &  \\
    & $\times (\lambda^{6}-1491\lambda^{4}+421155\lambda^{2}-12006225)$ & No \\
\hline
$13/2$ & $\lambda^{2}(\lambda^{6}-2184\lambda^{4}+1012752\lambda^{2}-74794752)^{2}$ & Yes  \\
\hline
$7$ & $-\lambda (\lambda^{2}-784)(\lambda^{4}-2296\lambda^{2}+353808)$ &  \\
    & $\times (\lambda^{8}-3108\lambda^{6}+2236710\lambda^{4}-328692196\lambda^{2}+3773030625)$ & No \\
\hline
$15/2$ & $(\lambda^{8}-4284\lambda^{6}+4488102\lambda^{4}-1062230652 \lambda^{2}+22347950625)^{2}$ & Yes  \\
\hline
$8$ & $-\lambda (\lambda^{4}+6624\lambda^{2}+1900800)(\lambda^{4}+16704\lambda^{2}+28753920)$ &  \\
    & $\times (\lambda^{8}+23184\lambda^{6}+138054240\lambda^{4}+204233529600\lambda^{2}+33886369440000)^{2}$ & No \\
\hline
$17/2$ & $\lambda^{2}(\lambda^{8}-7752\lambda^{6}+16263696\lambda^{4}-9531032320\lambda^{2}+995361177600)^{2}$ & Yes  \\
\hline
$9$ & $-\lambda (\lambda^{4}-7056\lambda^{2}+6441984)(\lambda^{4}-3096\lambda^{2}+668304)$ &  \\
    & $\times (\lambda^{10}-10197\lambda^{8}+29403594\lambda^{6}-25878927978\lambda^{4}+5213177173701\lambda^{2}-88322873900625)$ & No \\
\hline
$19/2$ & $(\lambda^{10}-13167\lambda^{8}+50640282\lambda^{6}-62764022286\lambda^{4}+19627235976789\lambda^{2}-584689432201875)^{2}$ & Yes  \\
\hline
$10$ & $-\lambda (\lambda^{4}-5456\lambda^{2}+3165184)(\lambda^{6}-11396 \lambda^{4}+20438704\lambda^{2}-2031480000)$ &  \\
    & $\times (\lambda^{10}-16797\lambda^{8}+84869994 \lambda^{6}-145160193178\lambda^{4}+68747106284901\lambda^{2}-3870591128105625)$ & No \\
\hline
$21/2$ & $\lambda^{2}(\lambda^{10}-21252\lambda^{8}+140008176\lambda^{6}-329460868800\lambda^{4}+241815611520000\lambda^{2}-33685691719680000)^{2}$ & Yes  \\
\hline
$11$ & $-\lambda (\lambda^{4}-8976\lambda^{2}+10644480)(\lambda^{6}-17556\lambda^{4}+55226160\lambda^{2}-15437822400)$ &  \\
    & $\times (\lambda^{12}-26598\lambda^{10}+225185103\lambda^{8}-712278892116\lambda^{6}$ &\\
    &$+768687668037135\lambda^{4}
    -202420859545362150\lambda^{2}4712996874211250625)$ & No \\
\hline\hline
\end{tabular}
\label{table:table1}
\end{table}
\end{widetext}
\begin{eqnarray}
\label{eq:WineSqueeze2}
\xi_{z}&=&\sqrt{4}\frac{\langle \left(\Delta J_{z}(t)\right)\rangle}{\left|\langle J_{x}(t)\rangle \right|},\nonumber\\
&=&\frac{2[7-3\cos4\sqrt{3}\chi t-(\sin6\chi t+\sqrt{3}\sin2\sqrt{3}\chi t)^{2}]^{1/2}}
{\left|\cos3\chi t(1+3\cos2\sqrt{3}\chi t)+\sqrt{3}\sin3\chi t\sin2\sqrt{3}\chi t\right|}.\nonumber\\
\end{eqnarray}
Squeezing occurs when the squeezing parameter is less than $1$. The parameter $\xi_{y}$ is plotted versus time in
Fig.~\ref{fig:P1}. There is no squeezing in the $y$ quadrature for the duration shown.
The parameter $\xi_{z}$ is plotted in Fig.~\ref{fig:P2}. Squeezing in the $z$ quadrature can be seen for two short 
intervals in the diagram.
\begin{figure}
\includegraphics[width=0.4\textwidth]{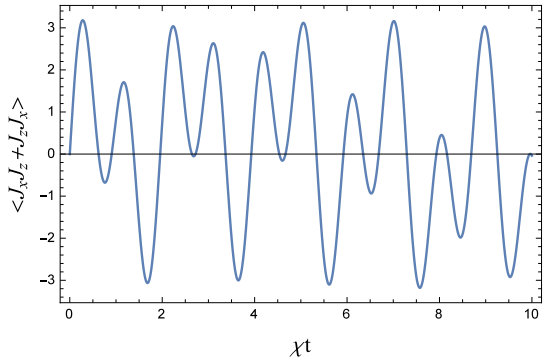}
\caption{(Color online) The correlation $\langle J_{x}J_{z}+J_{z}J_{x}\rangle$ [Eq.~(\ref{eq:corr})]
as a function of the dimensionless time $\chi t$ for $J=2$.}
\label{fig:P3}
\end{figure}

For completeness we mention that correlations between the various spin components, which are relevant 
to squeezing \cite{Ma2011}, can also be found analytically for this solvable model. For example, 
plotted in Fig.~\ref{fig:P3} 
is the quantity
\begin{eqnarray}
\label{eq:corr}
\langle J_{x}J_{z}+J_{z}J_{x}\rangle&=
&\frac{3}{2}\cos\sqrt{3}\chi t\nonumber\\
&&\times\left[\left(1-\sqrt{3}\right)\sin \left(3-\sqrt{3}\right)\chi t\right.\nonumber\\
&&+\left.\left(1+\sqrt{3}\right)\sin\left(3+\sqrt{3}\right)\chi t\right].\\
\nonumber
\end{eqnarray}

\section{Conclusion}
We have shown that the two-axis counter-twisting spin squeezing Hamiltonian can be solved analytically
for up to angular momentum $21/2.$ We have discussed the properties of the Hamiltonian that lead to
such a high degree of solvability. From our results the axis of optimum squeezing can be found readily.
Our methods can also be used to find useful quantities such as entanglement measures, in closed form. 
Future work will investigate the effects of decoherence on the solutions. We would like to thank K. 
Hazzard for stimulating discussions.

\section{Appendix}
In this Appendix we show that the anticommutation of Eq.(\ref{eq:AntiC}) implies the pairing of eigenvalues
of $H_{TA}$. Consider an eigenvector $\psi_{+}$ of $H_{TA}$ with an eigenvalue $\lambda$, i.e.
\begin{equation}
\label{eq:PlusH}
H_{TA}\psi_{+}=\lambda\psi_{+},
\end{equation}
Multiplying from the left by $e^{i\pi J_{y}}$ and using the anticommutation of Eq.~(\ref{eq:AntiC}),
we find the left-hand side of Eq.~(\ref{eq:PlusH}) reads
\begin{equation}
\label{eq:LHS}
e^{i\pi J_{y}}H_{TA}\psi_{+} = - H_{TA}e^{i\pi J_{y}}\psi_{+},
\end{equation}
while the right-hand side reads
\begin{equation}
\label{eq:RHS}
e^{i\pi J_{y}}(\lambda\psi_{+}) = \lambda(e^{i\pi J_{y}}\psi_{+}).
\end{equation}
Equating the right hand sides of Eqs.~(\ref{eq:LHS}) and (\ref{eq:RHS}), we arrive at
\begin{equation}
H_{TA}(e^{i\pi J_{y}}\psi_{+}) = -\lambda(e^{i\pi J_{y}}\psi_{+}),
\end{equation}
which implies that
\begin{equation}
\psi_{-} = e^{i\pi J_{y}}\psi_{+},
\end{equation}
is an eigenfunction of $H_{TA}$ with an eigenvalue of $-\lambda$. Thus the anticommutation
of the operator $e^{i\pi J_{y}}$ with $H_{TA}$ leads to the $\pm \lambda$ pairing of eigenvalues in
the spectrum of $H_{TA}$.

\end{document}